\documentclass{llncs}
\usepackage{graphicx}
\usepackage{subfigure}
\usepackage{amsmath,amsfonts}
  {\begin{list}{}{%
    \setlength{\labelwidth}{35pt}%
    \setlength{\leftmargin}%
                 {\labelwidth+\labelsep}}}%
  {\end{list}}

\newcommand{\ie}{i.e. }

\begin{document}
\title{Evolving Quantum Oracles with Hybrid Quantum-inspired Evolutionary Algorithm
\thanks{This work is partly supported
by the National Natural Science Foundation of China (No.60233010 and No.60496324), the
National Key Research and Development Program of China (No.2002CB312004), the Knowledge
Innovation Program of the Chinese Academy of Sciences, and MADIS of the Chinese Academy
of Sciences.}}
\author{Shengchao Ding\inst{1,3}, Zhi Jin\inst{1,2}, Qing Yang\inst{4}}
\institute{Institute of Computing Technology, Chinese Academy of Sciences \and Academy of
Mathematics and Systems Science, Chinese Academy of Sciences
\and Graduate University of the Chinese Academy of Sciences\\
Beijing 100080, China
\and School of Computer Science and Technology, South-Central
University for
Nationalities, Wuhan 430074, China\\
\email{dingshengchao@ict.ac.cn}}%
\maketitle

\begin{abstract}
Quantum oracles play key roles in the studies of quantum computation and quantum
information. But implementing quantum oracles efficiently with universal quantum gates is
a hard work. Motivated by genetic programming, this paper proposes a novel approach to
evolve quantum oracles with a hybrid quantum-inspired evolutionary algorithm. The
approach codes quantum circuits with numerical values and combines the cost and
correctness of quantum circuits into the fitness function. To speed up the calculation of
matrix multiplication in the evaluation of individuals, a fast algorithm of matrix
multiplication with Kronecker product is also presented. The experiments show the
validity and the effects of some parameters of the presented approach. And some
characteristics of the novel approach are discussed too.
\end{abstract}

\section{Introduction}
Quantum computation is a flourishing research area and it has been believed that quantum
computers hold a computational advantage over classical ones~\cite{00NielsenChuang}.
Generally, for simplification in the studies of quantum computation, a transformation or
even a quantum gate which isn't directly implemented physically, is treated as a black
box, \ie quantum oracle. 
However, one of the challenges implementing practical quantum computers is to design the
quantum oracle with quantum circuits made of available quantum
gates~\cite{89DeutschCiruit}. Mathematically, designing a quantum oracle can be
formulated as decomposing the expected unitary matrix to some smaller matrices which
correspond to the primary quantum gates. But the best known mathematic algorithms are not
very efficient~\cite{00NielsenChuang}. Moreover, the potential mechanics of quantum
computation is not well understood yet. So it's very difficult to develop heuristic
approaches designing quantum circuits at present. Finally, the designed quantum circuit
may be inefficient. Appropriate optimization techniques are required to reduce its cost,
such as rewriting rules based~\cite{02Iwama} or templates based~\cite{05Maslov}
techniques.

However, we could get some surprised results with genetic programming~\cite{92Koza}.
Genetic programming is a sort of robust optimization algorithm, which mimics natural
evolution that encodes the solution with chromosome, crossovers the old individuals
according to the fitness, mutates them with probability and obtains the new individuals
each generation over and over again. It is used to figure out complex optimization
problems and requires little prior knowledge of the problems. All the advantages of
genetic programming smooth out the above mentioned difficulties in designing quantum
oracles, thus it has been used to automatic designing quantum
circuits~\cite{98Williams,99Spector,00Yabuki,01Rubinstein,02Lukac,03Perkowski,05Reid}.

In this paper we develop a novel approach to evolve quantum oracle with a Hybrid
Quantum-inspired Evolutionary Algorithm (HQEA) which has been successfully applied in
numerical optimization and $0$-$1$ knapsack problems (see~\cite{06QYangMaster} for detail
introduction of HQEA). Taking the matrix corresponding to an oracle as input, the
presented approach can achieve designing and optimizing the quantum circuits at the same
time. This paper is organized as follows. In Section~\ref{sec:novelalgo}, the novel
approach evolving quantum oracles is presented. To speed up the evaluation of
individuals, a fast algorithm for matrix multiplication with Kronecker product is
presented in Section~\ref{sec:kronecker} too. The validity of the presented approach is
shown and the effects of some parameters of the approach are discussed in
Section~\ref{sec:experiments}. In the last section, some conclusions are drawn.

\section{Novel approach evolving quantum oracles}\label{sec:novelalgo}
The novel approach takes HQEA as optimization algorithm. To take advantage of HQEA, it is
necessary to encode quantum circuit with numerical value, which is different to the
symbol notations in the previous works. Here, the different quantum gates or the same
gates operating on different qubits are treated as different cases, which correspond to
different number. Although it is possible that several quantum gates are operated
parallelly, for simplification, it is assumed that only one quantum gate is applied on
some qubits each time.

\begin{figure}
\centering \includegraphics[width=0.66\textwidth]{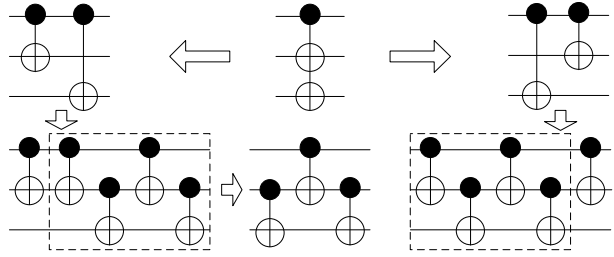} %
\caption{Replace non-adjacent $\mathrm{CNOT}$ with adjacent
$\mathrm{CNOT}$}\label{fig:cnotreplace}
\end{figure}

In the previous
works~\cite{02Iwama,05Maslov,98Williams,99Spector,00Yabuki,01Rubinstein,02Lukac,03Perkowski,05Reid},
the quantum gates operating on non-adjacent or multiple qubits are taken as primitive
gates. The commonly used non-adjacent quantum gates, for example, are generalized
$\mathrm{CNOT}$ in which a $\mathrm{NOT}$ operation on a qubit is controlled by a
non-adjacent one. In fact it's difficult to implement the quantum operations on distant
qubits. It's more reasonable to take some one-qubit and adjacent two-qubit operations as
primitive gates. One might argue that a realizable circuit could be obtained by replacing
the generalized $\mathrm{CNOT}$ with the equivalent circuit composed by adjacent
$\mathrm{CNOT}$. But it is likely to generate redundances, see Fig.~\ref{fig:cnotreplace}
as an illustration. In this paper, the available primitive gates include
Phase($\mathrm{S}$), $\pi/8$($\mathrm{T}$), Hadamard($\mathrm{H}$) and
$\mathrm{CNOT}$~\cite{00NielsenChuang}. But our approach is not confined to these gates.
Generally speaking, if $n_1$ one-qubit gates and $n_2$ adjacent two-qubit gates are
available, there are $N=n_1 m+2n_2(m-1)+1$ different cases on $m$ qubits (including the
quantum wire). So $k=\lceil \log_2N \rceil$ bits are required to encode one case. And if
a codon is decoded as integer $s$, it corresponds to the case indexed by $\lfloor
\frac{s\cdot N}{2^k} \rfloor$. An individual representing a quantum circuit is consisted
of $gk$ qubits where $g$ is the maximal number of allowable gates for the
circuit.\vspace{-14pt}
\vspace{-10pt}

Since even the same primitive gates have different costs depending on realization
technologies of quantum computers, the cost of the designed quantum circuits should be
considered. However, our approach is independent to the used cost function. In our work
we just assign the one-qubit gate costs one, two-qubit gate costs two, and the quantum
wire costs nothing. The cost of a circuit, $allcost$, is the total cost of the gates
presented in the circuit. Let $\mathrm{C}$ be the implemented circuit which corresponds
to the matrix $ \lambda(\mathrm{C}) = (I_{2^{m_{g}}}\otimes A_{g} \otimes I_{2^{n_{g}}})
\times \cdots \times (I_{2^{m_{1}}}\otimes A_{1} \otimes I_{2^{n_{1}}})$, where $m_i$ and
$n_i$ $(i=1,\cdots,g)$ are respectively the numbers of qubits before and after the gate
$A_i$. If $G$ be the goal matrix the circuit should implement, the correctness of
$\mathrm{C}$ is defined by~\cite{05Reid} such as
        \begin{equation}
            correctness(\mathrm{C}) = \frac{|tr(G^{\dag}\lambda(\mathrm{C}))|}{2^m}
        \end{equation}

The fitness function takes $allcost$ into account as well as the correctness of the
circuit. Evolving quantum oracles is time-consuming when the scale of problem is very
big. Sometimes we could be satisfied with a non-optimal circuit whose cost is less than
some value. Such a value is called the satisfying cost ($satcost$). But it's more
important to obtain a correct quantum circuit. So a tradeoff between the cost and
correctness should be taken. We take two thresholds $award$ and $punish$, and then the
fitness function is defined as
\begin{equation}
fitness(\mathrm{C})=award\times(allcost-satcost) + punish\times
(1-correctness(\mathrm{C}))
\end{equation}

As an evolutionary algorithm, the termination condition of the proposed approach is
meeting the satisfying cost as well as fulfilling the correctness, or evolving allowable
generation.

\section{Fast matrix multiplication with Kronecker product}\label{sec:kronecker}
Lots of matrix multiplications are required in the process of evaluating individuals. It
is well known that we should perform $O(n^3)$ multiplications when two $n\times n$
matrices are multiplied naively. Although the best algorithm currently known has an
asymptotic complexity of $O(n^{2.376})$~\cite{94Horn}, some improvement is still possible
for the matrix multiplication with Kronecker product which is required in this paper.
Firstly, we put up some conventions.
\begin{itemize}
    \item Kronecker product, matrix multiplication and scalar product are denoted as $\otimes$,
    $\times$ and $\cdot$ respectively. The priority of $\cdot$ is higher
    than others and $\times$ takes on the lowest priority.
    \item $A_n^{(k)}$ denotes arbitrary $n\times n$ matrix $A$ treated as $k\times k$
    blocks, each of which is a $\frac{n}{k}\times \frac{n}{k}$ matrix. Specially, $A_n$ means $A_n^{(1)}$
    and $\mathbf{1}_n$ denotes the $n\times n$ identity matrix.
    \item 
    $B(i,j)_{nk}$ denotes the block located at the $i$-th row and the $j$-th column in $B_{mnk}^{(m)}$. In
    particular, $A(i,j)_1$ denotes an element of $A$, for short, $a_{ij}$.
\end{itemize}

The trick is based on the block multiplication and the calculation is shown as follows.
\begin{equation}\label{eqn:kronproduct}
\begin{array}{ccl}
F & = & (\mathbf{1}_{m} \otimes A_{n} \otimes \mathbf{1}_{k}) \times B_{mnk} \\
& = & \left(
\begin{array}{ccc}
A_n & & \\
& \ddots & \\
& & A_n \\
\end{array} \right)_{mn}^{(m)} \otimes \mathbf{1}_k \times B_{mnk} \\
& = & \left( \begin{array}{ccc} A_n\otimes \mathbf{1}_k & & \\
& \ddots & \\
& & A_n \otimes \mathbf{1}_k \\
\end{array} \right)_{mnk}^{(m)} \times \left(
\begin{array}{ccc}
\cdots & \cdots & \cdots \\
\vdots & B(i,j)_{nk} & \vdots\\
\cdots & \cdots & \cdots \\
\end{array}
\right)_{mnk}^{(m)}\\
& = & \left( \begin{array}{ccc} \cdots & \cdots & \cdots
\\
\vdots & A_n\otimes \mathbf{1}_k\times B(i,j)_{nk} & \vdots \\
\cdots & \cdots & \cdots \\
\end{array}\right)_{mnk}^{(m)}
\end{array}
\end{equation}
Let $A_n\otimes\mathbf{1}_k\times B(i,j)_{nk} \equiv D = F(i,j)_{nk}$, ($i,j=1\ldots m$),
then
\begin{equation}
F(i,j)_{nk} =  \left(
\begin{array}{ccc}
\cdots & \cdots & \cdots \\
\cdots & a_{pq}\cdot \mathbf{1}_k & \cdots \\
\cdots & \cdots & \cdots \\
\end{array}
\right)_{nk}^{(n)} \times \left(
\begin{array}{ccc}
\cdots & \cdots & \cdots \\
\cdots & B(i,j)(p,q)_k & \cdots \\
\cdots & \cdots & \cdots \\
\end{array}
\right)_{nk}^{(n)}
\end{equation}
\begin{equation}
D(p,q)_k = \sum_{l=1}^{n}\left( a_{pl}\cdot
B(i,j)_{nk}(l,q)_{k}\right) (p,q=1\ldots n)
\end{equation}

It's very clear that obtaining $D(p,q)$ requires $O(nk^2)$ multiplications, and then
$O(n^3k^2)$ multiplications to obtain $D$ and lastly $O(m^2n^3k^2)$ multiplications to
obtain $F$. So the fast algorithm speeds up
$\frac{(mnk)^{2.376}}{m^2n^3k^2}={(mk)^{0.376}}/{n^{0.624}}$ to the original best
algorithm. Simple algebra shows that our fast algorithm exceeds the best traditional
algorithm when the number of qubits before the gate, $M=\log_2{m}$, the number of qubits
after the gate, $K=\log_2{k}$, and the number of qubits the gate processes,
$N=\log_2{n}$, satisfy $M+K>1.66N$.

\section{Experiments and discussions}\label{sec:experiments}
\begin{figure}
\centering \subfigure[Optimal circuit]{\label{fig:swap1}
\qquad\qquad\includegraphics[width=0.17\textwidth]{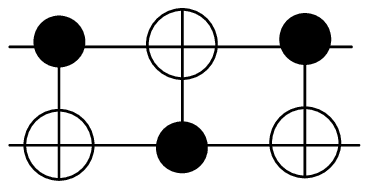}\qquad\qquad}
\subfigure[Non-optimal circuit]{\label{fig:swap2}
\qquad\qquad\includegraphics[width=0.29\textwidth]{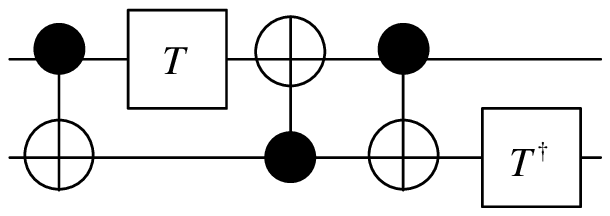}\qquad\qquad}
\caption{Equivalent circuits of $\mathrm{SWAP}$}
\end{figure}

The presented approach is not confined to circuit design or circuit optimization, and the
key factor is the satisfying cost. When this constraint is loose corresponding to a
bigger satisfying cost, the algorithm performs mainly as an automatic designer which aims
to discover a circuit implementing the desirable unitary transformation, and regardless
whether the result is optimal. If this constraint is tight corresponding to a smaller
satisfying cost, the algorithm not only tries to dig out a circuit functioning as
desired, but also to reduce the cost of the circuit. As an instance, the optimal circuit
of oracle $\mathrm{SWAP}$ is $\mathrm{CNOT}*\mathrm{CNOT2}*\mathrm{CNOT}$
($\mathrm{CNOT2}$ denotes the $\mathrm{CNOT}$ taking the latter qubit as control bit and
the former qubit as controlled bit) such as Fig.~\ref{fig:swap1} which costs $6$, but if
we assign $satcost$ as $8$, another circuit could be obtained by our approach, such as
Fig.~\ref{fig:swap2}.\vspace{-5pt}
\vspace{-25pt}
\begin{table}
\begin{minipage}{\textwidth}
\centering \caption{Comparison with various satisfying costs}\label{tbl:satcost}
\begin{tabular}{@{\ }cp{4em}p{5em}p{70pt}p{5em}ccc@{\ }}
\hline
 Goal oracle & Optimal cost & Satisfying cost & Maximal number of allowable gates & Maximal generation & $AS$ & $ST$ &
 $OT$\footnote{$AS$ = average generation of success, $ST$ = times of success in all tests, $OT$ = times
of getting optimal results in all tests.}\\ \hline
 \emph{entangle2} & 3 & 4 &  6  & 100& 86.1  &     20 & 4 \\
& &6  & 6 &  100& 14.8 &    20 & 0 \\
\emph{entangle3} & 5 & 6  & 8 &200& 141.7 &  20& 10 \\
&&8&   8&   200& 48.65&    20 & 1 \\
\emph{controlled-$\mathrm{S}$} & 7 & 8 &  8 &  500& 111.5 &   20 &3 \\
&&10&  8  & 500 &62.5  &20& 1\\
\hline\\
\end{tabular}
\end{minipage}\vspace{-10pt}
\end{table}

\begin{figure}
\centering \subfigure[Optimal one(cost 3)]{ \label{fig:entangle2-1}\quad \qquad
\includegraphics[width=0.17\textwidth]{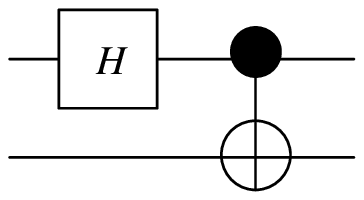}\quad \qquad} \subfigure[One
of non-optimal circuits(cost
5)]{\label{fig:entangle2-2}\quad\qquad\includegraphics[width=0.32\textwidth]{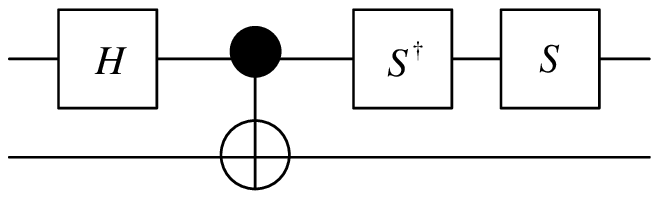}\qquad\quad}
\caption{Circuits entangling two qubits}\label{fig:entangle2}%
\vspace{8pt}%
\centering \subfigure[Optimal one(cost 5)]{
\label{fig:entangle3-1}\quad\includegraphics[width=0.23\textwidth]{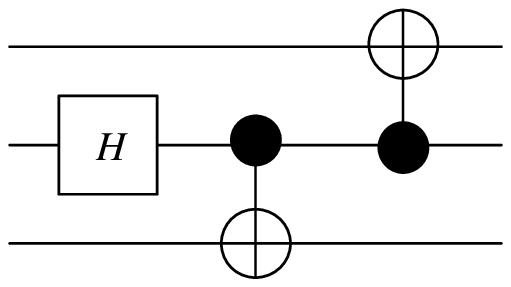}\quad }
\quad \subfigure[One of non-optimal circuits(cost
7)]{\label{fig:entangle3-2}\quad\qquad\includegraphics[width=0.3\textwidth]{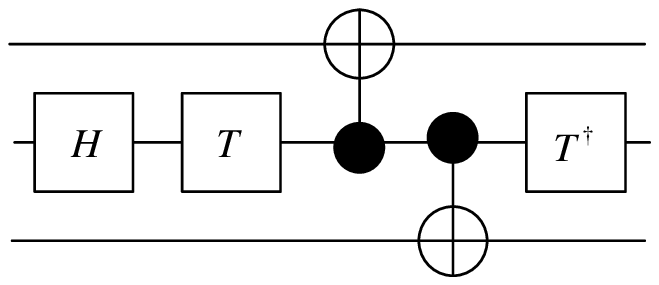}\quad\qquad}
\caption{Circuits entangling three qubits}\label{fig:entangle3} %
\vspace{8pt}%
\centering
\subfigure[Optimal one(cost 7)]{
\label{fig:cS-1}\includegraphics[width=0.32\textwidth]{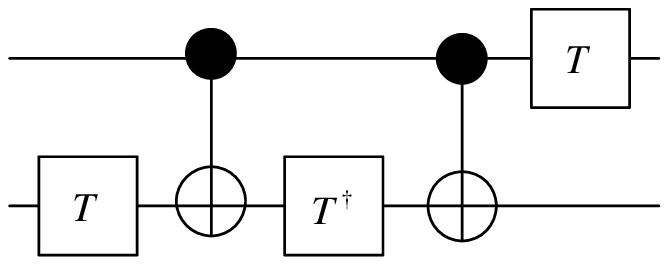} }%
\ \subfigure[One of non-optimal circuits(cost 8)]{
\label{fig:cS-2}\includegraphics[width=0.29\textwidth]{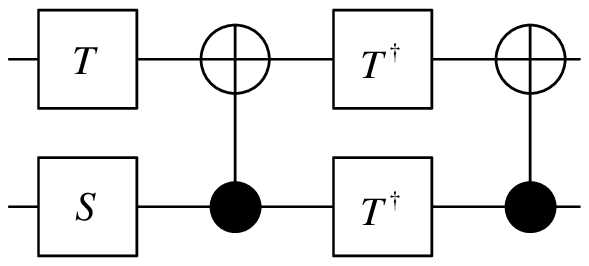}}%
\ \subfigure[One of non-optimal circuits(cost 9)]{
\label{fig:cS-8}\includegraphics[width=0.29\textwidth]{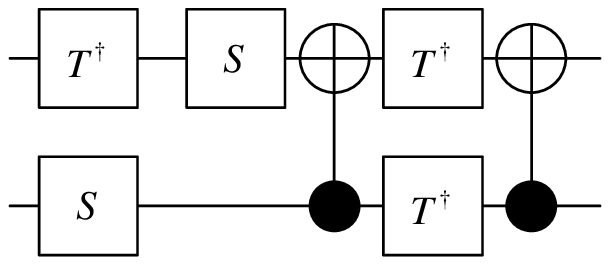} }%
\caption{Circuits implementing controlled-phase}\label{fig:cS}
\end{figure}

To study the affection of satisfying cost on evolving results, we apply novel algorithm
on the oracle entangling two qubits, denoted by {\em entangle2} as
Fig.~\ref{fig:entangle2}, the oracle entangling three qubits, denoted by {\em entangle3}
as Fig.~\ref{fig:entangle3} and the oracle implementing controlled-phase, denoted by {\em
controlled-$\mathrm{S}$} as Fig.~\ref{fig:cS}. Each case is tested $20$ times and in all
the test, the number of quantum chromosome is $20$, measurement times of each quantum
chromosome is $10$. The comparison results are shown in Table~\ref{tbl:satcost}. In the
experiment, all the tests obtain the circuits satisfying condition successfully. However,
obtaining the circuits meeting more rigorous constraints needs more evolving time.
Additionally, it can be found that there are many equivalent circuits implementing the
same oracle, although some of them cost differently. Some of these circuits are
illustrated in Fig.~\ref{fig:entangle2},\ref{fig:entangle3},\ref{fig:cS}.
\vspace{-20pt}

To effectively discover  desired quantum circuits for different cases, it is useful to
adopt the appropriate reward-punish factor in the fitness function.
Table~\ref{tbl:factors} shows the evolutionary results of oracle \emph{entangle2} with
different reward-punish factors, where each case is tested $20$ times. It is found that
bigger punishment to the error of circuits is required to get the correct circuits, with
the same $satcost$, $g$ and other parameters. Another fact is that with the same $g$ and
other parameters, to get the correct circuits bigger $punish$ is required for bigger
$satcost$.
\begin{table}
\begin{minipage}{\textwidth}
\centering \caption{Comparison with different reward-punish
factors($reward=1$)}\label{tbl:factors}
\begin{tabular}{@{\ }p{5em}p{70pt}p{5em}ccc@{\ }}
\hline Satisfying cost & Maximal number of allowable gates & Maximal generation & $ST$ &
$AS$\footnote{$ST$ = times of success in all tests, $AS$ = average generation of
success.} & $punish$
\\ \hline
6&   6  & 100& 0 &  0 &  1\\
 &   &  100& 20  &16.9 &   5\\
 &    & 100 &20 & 10.65 &  20\\
 &    & 100 &20 & 12.8  &  100\\
 &    & 100 &20 & 15.55  & 1000\\
8 &  8 &  200& 0 &  0 &  1\\
 &    & 200& 0 &  0  & 5\\
 &    & 200 &20 & 32 & 20\\
 &     &200 &20 & 94  &100\\
&     & 200& 20 & 62.65 &  1000\\
10 & 8  & 500& 0 &  0  & 1\\
 &    &500 &0  & 0 &  5\\
&   &  500& 0  & 0 &  20\\
&    & 500& 20 & 40.45 &  100\\
 &   & 500 &20 & 31.1 &   1000 \\ \hline
\end{tabular}
\end{minipage}\vspace{-8pt}
\end{table}

When comparing our work with
others~\cite{02Iwama,05Maslov,98Williams,99Spector,00Yabuki,01Rubinstein,02Lukac,03Perkowski,05Reid},
we observe the following aspects:
\begin{enumerate}
    \item Problems considered: Our approach evolves quantum circuits taking the desired unitary matrix as
    input, while some other works are based on the description of an oracle and evolve quantum circuits by comparing
     the outputs of the quantum oracle on random inputs with desirable ones~\cite{00Yabuki}.
     In addition, the simplification of known quantum circuits are considered
     by~\cite{02Iwama,05Maslov}. Notably, only reversible quantum oracles are considered in our
     work. Thus measurements, not like the cases in~\cite{99Spector,00Yabuki}, are not permitted.
    \item Primary gates set: As has been stated, only two-qubit gates on adjacent qubits are
    available in our work while non-adjacent two-qubit or even multi-qubit gates are taken as primary gates in other
    works,  although it can simplify the problem of quantum circuit design. Of course, our approach is not confined to any special
    quantum gates set.
    \item Circuit representations: While all the previous works use symbolic representations, our approach encodes
    the quantum circuits with numerical values. In despite of the apparent difference within them, all of them are
    equivalent. But it is more natural to apply evolutionary operators on the individuals coded by our means.
    \item Fitness function: The cost of designed quantum circuits is ignored
    in~\cite{98Williams,99Spector,00Yabuki,01Rubinstein}. Moreover, the cost and
    correctness are dividually considered in~\cite{05Reid}. To reflect the fitness of evolved quantum circuits more accurately and
    expediently, our approach combines them together by simple reward-punish factors.
    \item Application: Our approach can be adapted to both circuit designing and circuit
    optimization. When applied to the later, more rigorous conditions should be assigned,
    such as the smaller satisfying cost and maximal gates number.
    \item Common challenge: The bottleneck in designing quantum circuits with evolutionary algorithms is the individual
    evaluation, \ie the matrix multiplications in the computing of fitness which have potentially exponential space
    increment and speed slow down. The intractable problem results from the argument that the quantum system can not be
    effectively classically simulated.
\end{enumerate}

\section{Conclusions}
Genetic programming appears to be useful in designing quantum circuits. We propose how to
evolve quantum oracle with a hybrid quantum-inspired evolutionary algorithm. With our
approach no additional knowledge is required to design an optimized quantum oracle as
expected. Especially, we design a novel approach to represent quantum circuits with
numerical values in the evolutionary algorithm. A faster algorithm for matrix
multiplication with Kronecker product is presented too. It speeds up the evaluation of
individuals very much. Obviously, the numerical representation and fast algorithm of
matrix multiplication are not unique to our approach, but adaptable to other evolutionary
quantum programming algorithms or hierarchical approaches. By assigning different
parameters, the novel approach could be inclined to designing a quantum circuit or
optimizing it. Our approach provides insights into quickly evolving quantum oracles. A
possible improvement to the approach may be encoding the quantum circuits with variable
length.

\bibliography{thebib}
\end{document}